\documentclass[11pt]{article}

\usepackage{graphicx,epsfig}
\usepackage{multicol}
\usepackage{amsmath,amssymb,cite,color,hyperref}
\newcommand{\be}{\begin{equation}}
\newcommand{\ee}{\end{equation}}
\newcommand{\bea}{\begin{eqnarray}}
\newcommand{\eea}{\end{eqnarray}}
\newcommand{\bi}{\begin{itemize}}
\newcommand{\ei}{\end{itemize}}
\newcommand{\ben}{\begin{enumerate}}
\newcommand{\een}{\end{enumerate}}

\def\frac#1#2{{{#1}\over {#2}}}
\def\gsim{\mathrel{\rlap{\lower4pt\hbox{\hskip1pt$\sim$}}
    \raise1pt\hbox{$>$}}}         
\def\lsim{\mathrel{\rlap{\lower4pt\hbox{\hskip1pt$\sim$}}
    \raise1pt\hbox{$<$}}}         

\newcommand{\draft}[1]{}

\definecolor{grey}{rgb}{0.5,0.5,0.5}

\usepackage{cite}
\usepackage{hyperref}
\usepackage{url}

\begin{document}

\begin{flushright}
IFUM-1021-FT\\
\end{flushright}

\begin{center}
{\large\bf Do we need N$^3$LO Parton Distributions?}
\vspace{0.6cm}

Stefano Forte, Andrea Isgr\`o and Gherardo Vita\\ 

\vspace{.3cm}
Dipartimento di Fisica, Universit\`a di Milano and
INFN, Sezione di Milano,\\ Via Celoria 16, I-20133 Milano, Italy\\
\end{center}   

\vspace{0.2cm}

\begin{center}
{\bf \large Abstract}
\end{center}
We discuss the uncertainty on processes computed using next-to-next-to
leading (NNLO) parton distributions (PDFs) due to the neglect of
higher order perturbative corrections in the PDF determination, in the
specific case of Higgs production in gluon fusion. By studying the
behaviour of the perturbative series for this process, we show that
this uncertainty is negligible in comparison to the theoretical uncertainty on the
matrix element. We then take this as a case study for the use of the
Cacciari-Houdeau method for the estimate of theoretical uncertainties,
and show that the method provides an effective way of treating
theoretical uncertainties on the matrtix element and the PDF on the
same footing. We briefly discuss the possible generalization of these
results to other processes, and in particular top production.

\clearpage
Gluon fusion, the dominant Higgs production channel at the LHC, has a
slowly convergent  expansion in  perturbative QCD: the inclusive cross
section is currently known up to next-to-next-to-leading order
(NNLO)~\cite{higgslo,higgsnlo,higgsnnlo},
and a recent approximate determination of the N$^3$LO result has been
presented~\cite{Ball:2013bra}, while rapid progress on the exact
computation has been reported~\cite{higgsn3lo}.

With N$^3$LO results around the corner, it is natural to ask whether
these will be of any use, given that fully consistent 
N$^3$LO parton distributions
(PDFs) are not likely to be available any time soon, essentially
because the determination of N$^3$LO anomalous dimensions would
require a fourth-order computation, for instance of deep-inelastic
structure functions, or Wilson coefficients. Clearly, this question is
related to the more general issue of theoretical uncertainties on
PDFs: current PDF uncertainties~\cite{Forte:2013wc} only reproduce the uncertainty in the
underlying data, and of the procedure used to propagate it onto PDFs,
but not that related to missing higher-order corrections in the theory
used for PDF determination. Henceforth in this paper we will call
`theoretical uncertainty' the uncertainty due to the fixed-order
truncation of the perturbative expansion, sometimes~\cite{David:2013gaa} also called
missing higher-order uncertainty, or MHOU.

Here we address this set of issues in the specific context of Higgs
production in gluon fusion. We use the dependence on the
perturbative order of the prediction for this process as either the
PDF or the matrix element are taken at different orders as an
estimate the theoretical uncertainty on either. We then  address
the more general issue of how one may estimate theoretical
uncertainties on PDFs and matrix elements, specifically by
using the approach of
Cacciari and Houdeau~\cite{Cacciari:2011ze}.

\begin{figure}[t]
\centering
\begin{center}
\includegraphics[width=0.6\textwidth]{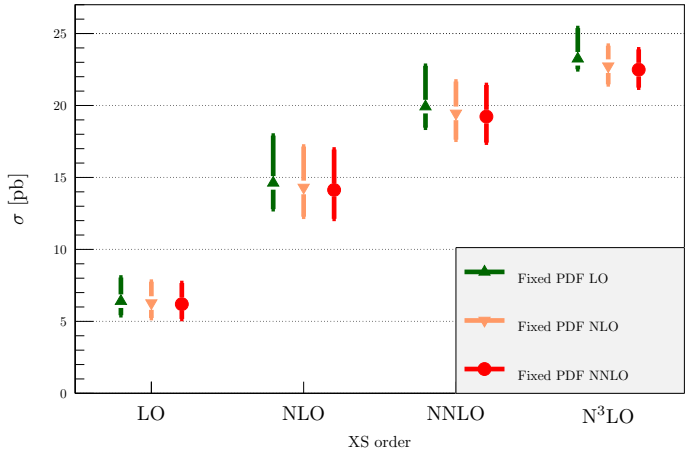} 
\end{center}
\caption{The cross section for Higgs production in gluon fusion,
  computed varying the perturbative order of the matrix element. The
  label on the $x$-axis denotes the order of the matrix element, while
in each case the three points from left to right are obtained
respectively using LO, NLO, and NNLO PDFs. The uncertainties are obtained
varying the renormalization scale by a factor two about $\mu_R=m_H$. The
N$^3$LO result is the approximation of Ref.~\cite{Ball:2013bra}.}
\label{fig:orderdep}
\end{figure}
We first compute the cross-section using the {\tt ggHiggs}
code~\cite{Ball:2013bra,gghiggs}, with default settings~\footnote{We
  have checked that similar results are obtained using {\tt
    ihixs}~\cite{Anastasiou:2011pi} version
  1.3.3. Note that previous versions of this code instead disagreed
  with {\tt ggHiggs},
  because of bugs affecting the top mass dependence at NLO and the
  factorization scale. Minor differences persist in the top mass
  dependence, but  {\tt ggHiggs} fully agrees with the non-public
  code of Ref.~\cite{Bonciani:2007ex}.}. Results are shown in
Figure~\ref{fig:orderdep}, where we show the cross section evaluated at
increasingly high perturbative order (henceforth loosely referred to as the
``order of the matrix element''), also including 
the approximate N$^3$LO from Ref.~\cite{Ball:2013bra}, using in each
case LO, NLO or NNLO PDFs (henceforth referred to as the ``order of
the PDF'').  We use NNPDF2.3
PDFs~\cite{Ball:2012cx} (with NNPDF2.1~LO~\cite{Ball:2011uy} as LO
set~\cite{Carrazza:2013axa}). What is shown here is the total
cross-section at the hadronic level, obtained summing over all parton
subchannels, except at N$^3$LO, where only the gluon-gluon channel is
included in the estimate of Ref.~\cite{Ball:2013bra}.

We assume $\alpha_S(M_Z)=0.119$ in all
cases, as we are interested in studying the behaviour of the
perturbative series for a fixed value of the coupling constant.
 The
uncertainty bars in Figure~\ref{fig:orderdep}  are all obtained by varying the renormalization
scale in the range $m_H/2\le \mu_R \le 2 m_H$ (the choice  $\mu_R=m_H/2$
as central scale 
is sometimes advocated instead~\cite{Anastasiou:2011pi}, as it leads to faster convergence of the perturbative
expansion: this choice would not change our conclusions). 
The variation of the renormalization scale should provide an estimate
of the missing higher-order corrections to the matrix element when the
PDF is kept fixed~\footnote{For this process, the dependence of results on the factorization scale is entirely
negligible even at LO~\cite{Ball:2013bra}.}, though, as well known, for
this process it substantially underestimates them.

Be that as it may, it is clear that the dependence of the result on
the order of the matrix element is much stronger than the dependence
on
the order of the
PDF: on the scale of the variation of the matrix element, results
obtained when the order of the PDF is varied are almost identical
(especially beyond LO). Hence we could conclude here our brief
investigation, having answered in the negative the question which is
asked in the title: at least as far as Higgs in gluon fusion is
concerned, based on the behaviour of the perturbative expansion at
known orders, it is very likely that using NNLO PDFs in the N$^3$LO
computation would lead to results which are essentially
indistinguishable from those consistently obtained using N$^3$LO PDFs
at N$^3$LO.

However, it is worth elaborating a little more on our
result. Specifically, it would be desirable to be able to provide a
quantitive estimate of the theoretical uncertainty on the PDF, as well
as of the combined theoretical uncertainty on the hadron-level process
due to both the matrix element and the PDF. 
In principle, it is possible to use scale variation in order to
determine the uncertainty on the PDF, too: it is, however, quite
cumbersome in practice as it
requires keeping track of the scale variation during the PDF
fitting~\cite{Olness:2009qd}. Indeed, to the best of our knowledge,
it has never been done for any
of the available PDF sets. Also, correlations between the behaviour of
different processes upon scale variation would have to be kept into
account: correlations between processes used for PDF determination
among each other would be needed in order to determine
the uncertainty on the PDF, and correlations between them and the
processes for which a
prediction is sought would be required in order to be able to combine
uncertainties on 
the PDF and the matrix element.

On the other hand, it is clear by looking at
Figure~\ref{fig:orderdep}  that a good deal of information is
contained in the behaviour of the perturbative expansion itself:
it is then natural to try to
systematically  provide a determination of the uncertainty bar
based on previous orders, rather than on scale variation. This has the further
advantage that the theoretical uncertainty due to the matrix
element and the PDF could then be treated on the same footing, and
easily combined.

\begin{figure}[t]
\centering
\begin{center}
\includegraphics[width=0.6\textwidth]{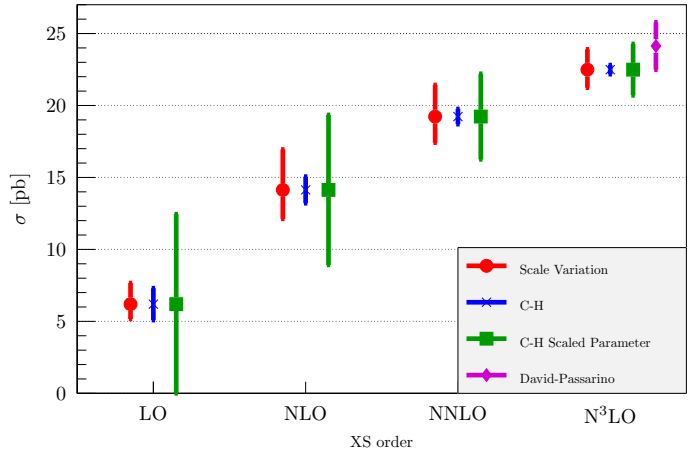} 
\end{center}
\caption{The cross section for Higgs production in gluon fusion
  computed with NNLO PDFs at increasing perturbative orders. At each
  order the uncertainty is shown  
as determined (from left to right) using scale variation
  (red circles, same as Fig.~\ref{fig:orderdep}), the Cacciari-Houdeau
method (blue crosses), and
  the same method  but with rescaled parameter (see text, green squares); at
  N$^3$LO the Passarino-David uncertainty is also shown (see text,
  purple diamonds).}
\label{fig:methcomp}
\end{figure}
A methodology to do so has been suggested
Ref.~\cite{Cacciari:2011ze} (Cacciari-Houdeau method, henceforth), based on assuming a prior distribution
for the coefficients of the perturbative expansion, and then using
Bayesian arguments to determine a confidence interval for the unknown
coefficients based on the behaviour of the known ones. In
Figure~\ref{fig:methcomp} we compare the theoretical uncertainty on
the matrix element  computed by scale variation, already shown in
Fig.~\ref{fig:orderdep} (corresponding to points shown in the plot as
red circles) to that obtained using the Cacciari-Houdeau method,
i.e. essentially Eq.~(85) of Ref.~\cite{Cacciari:2011ze} (shown as blue
crosses). In all cases, the PDF is kept fixed to the NNLO set. It
should be noticed that this is a very simple-minded application of
the ideas of Ref.~\cite{Cacciari:2011ze}: in particular, we do not
distinguish between different partonic subchannels, which could be in
principle characterized by different perturbative expansion, and 
we study the perturbative behaviour of the total hadronic cross section
rather than, for instance, the perturbative behaviour
of the differential partonic cross-section.

The result of Fig.~\ref{fig:methcomp}  is clearly not very
satisfactory: at each order, the uncertainty bar, rather than being of the
same order as  the shift when going to the next order, is much smaller
than it, and also
rather smaller than the scale uncertainty, which already
underestimated this shift. This could of course be due to the crude
approximations we are making, as discussed above. 

However, there is a  more
fundamental consideration. Namely, the approach of
Ref.~\cite{Cacciari:2011ze} is based on the assumption that the
perturbative expansion coefficients whose behavior is being studied
are 
roughly all of the same
order --- at least at low orders, well below the point 
where the perturbative series, which
is at best asymptotic, starts diverging. The result shown in Fig.~\ref{fig:methcomp} has been obtained
by writing the cross-section as a series  
\begin{equation}
\sigma=\alpha_S^2\left(\sigma_0+\alpha_S \sigma_1+\alpha_S^2 \sigma_2+
\alpha_S^3 \sigma_3+\dots\right)
\label{pertexp}
\end{equation}
and identifying the expansion coefficients $\sigma_i$ with the
coefficients $c_i$ as given in Eq.~(85) of
Ref.~\cite{Cacciari:2011ze}. However, it turns out that the
coefficients $\sigma_n$ thus defined 
rapidly grow with the perturbative order. On
the other hand, it is clear that (as already
pointed out in Ref.~\cite{Cacciari:2011ze}) another
choice of expansion parameter would generally
lead to a different behaviour. Indeed,
the natural expansion parameter, even in the simplest cases,
usually differs by the strong coupling by a (possibly large) factor: it might be
given, for
instance, by $\frac{\alpha_S}{4\pi}$ or $C_A\alpha_S$.

Lacking an analytic knowledge which may motivate a choice of the
expansion parameter (especially in view of our very simple-minded
approach), we  rewrite Eq.~(\ref{pertexp}) by rescaling
$\alpha_S$ by a real parameter $\lambda$:
\begin{eqnarray}
&&\sigma=\alpha_S^2\sigma_0\left(1+\bar\alpha_S c^\lambda_1+\bar \alpha_S^2 c^\lambda_2+
\bar\alpha_S^3  c^\lambda_3+\dots\right)\\
&&\quad \bar\alpha_S\equiv\lambda \alpha_S.
\label{pertexpa}
\end{eqnarray}
The approach of Ref.~\cite{Cacciari:2011ze}) is then applicable if
there exists a value of $\lambda$ such that the rescaled coefficients
$c^\lambda_i$ are all of comparable order.

We then simultaneusly test for the applicability of the method, and 
determine the optimal value of $\lambda$, by letting
$c_n^\lambda=\kappa$ and then performing a two-parameter fit of
$\lambda$ and $\kappa$ to the three known coefficients (including the
approximate N$^3$LO result of Ref.~\cite{Ball:2013bra}). We get an
almost perfect fit ($\chi^2$ below 1\%), and a
best-fit value of $\lambda=5.6$, with NNLO PDFs and $\mu_R=m_H$. The
good quality of the fit means that the rescaled coefficients are
indeed all of the same order, thus justifying the use of the
method, but  the large rescaling which is required explains the
failure of the method before rescaling.
Note that the rescaled expansion
parameter is large, but still  smaller than one, as one expects for a
slowly convergent series.
 
We
have checked that the best-fit $\lambda$ is quite stable upon variations of the
procedure. In particular it varies by a few percent if we change the order
    of the PDF, or if we decide to also include the leading-order coefficient
    in the fit 
(i.e.  if we fit directly the coefficients $\sigma_n$ of
Eq.~(\ref{pertexp}) as $\sigma_n=\kappa \lambda^n$): this latter
choice leads to a
significantly worse $\chi^2$,  but with
essentially the
same $\lambda$. If  we change the renormalization 
scale to $\mu_R=m_H/2$ the optimal $\lambda$ decreases by
about 20\%, to $\lambda=4.3$, while the fit quality
deteriorates to $\chi^2\sim1.1$, still justifying the use of the
method, given
that the equality of the coefficients is only expected to be approximate.\footnote{It is
  amusing to note that if one studies the fit quality as a function of
    $\mu_R$, the optimal fit turns out to have a very sharp minimum at
  $\mu_R=m_H$, where the fit is almost
  perfect ($\chi^2$ of order of $10^{-3}$). Otherwise stated, imposing equality of the
  coefficients $c_i^\lambda$ would determine
  $\frac{\mu_R}{m_H}=1\pm0.1$. Note however that this (presumably
  accidental) result relies on the
  approximate N$^3$LO value of $c_3$ of Ref.~\cite{Ball:2013bra}.}
The fact that the rescaling is somewhat smaller is an interesting feature of
the method: it shows that the perturbative expansion converges
somewhat faster with this choice of renormalization scale, as it is
known to be in fact the case.

Armed with the knowledge of the necessary rescaling $\lambda=5.6$, we 
recompute the Cacciari-Houdeau uncertainty using the
rescaled parameter $\bar\alpha_s$. 
The result is also shown in Fig.~\ref{fig:methcomp} (green
squares). It is clear that now the result provides a rather reasonable
estimate of the theoretical uncertainty, which, up to NNLO, 
turns out to be of the
same order as the observed perturbative shift at each order, and thus
in particular it reflects the theoretical uncertainty better than
scale variation. At N$^3$LO, scale variation and Cacciari-Houdeau lead
to similar answers. For
comparison, in Fig.~\ref{fig:methcomp} we also show (purple diamonds)
the uncertainty on the N$^3$LO result estimated according to the
method of Ref.~\cite{David:2013gaa}. In this reference, the
theoretical uncertainty is determined by assuming that the
perturbative series is an asymptotic series which is summed using
various techniques (such as Borel summation). For a same-sign
series the uncertainty band is taken to be at any given order as the
interval between the known result up to that order, and the upper
(more in general, the extreme) all-order asymptotic sum  --- so the lower
edge of the band coincides with the N$^3$LO central value, by
construction. Interestingly, the size of the uncertainty band on the
N$^3$LO result found using the method of
Ref.~\cite{David:2013gaa} is very close to that from Cacciari-Houdeau
(which, at this order, is also similar to scale variation as already mentioned).

\begin{figure}[t]
\begin{center}
\includegraphics[width=0.6\textwidth]{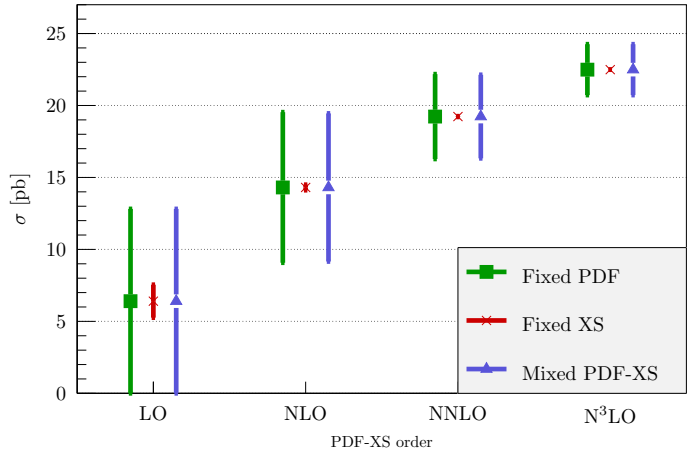} 
\end{center}
\caption{
Comparison of the theoretical uncertainty due to the matrix
  element (green squares), to the  PDF (red crosses), or both (violet
  triangles).  The cross section is
  computed at each  order using
  consistently the corresponding PDFs. All uncertainties are determined using the
  Cacciari-Houdeau method (see text for details).}
\label{fig:finres}
\end{figure}
We now finally turn to a determination of the theoretical uncertainty due to
either the PDF, or the matrix element, or both. Results are shown in
Figure~\ref{fig:finres}, for the hadronic cross-section computed at
each order using consistently PDFs at the corresponding order (LO
matrix element with LO PDFs and so on). All uncertainties are now computed
using 
the Cacciari-Houdeau method. In order to determine the uncertainty on
the matrix element as the PDF is kept fixed, we have used the rescaled
method as discussed above, with the PDF kept fixed either to its LO,
NLO, or NNLO value: we then show at LO the uncertainty on the matrix
element when the PDF is kept
fixed at LO, at NLO the uncertainty  when it is kept fixed at NLO and
so on. In order to determine the uncertainty due to the PDF, no
rescaling turns out to be necessary, so we use the method with $\alpha_S$
taken as expansion parameter. Finally, the combined uncertainty is
determined applying the rescaled Cacciari-Houdeau method to the series
at the hadronic level in which the order of the PDF and the matrix
element are varied simultaneously (with the NNLO PDF used also at
N$^3$LO); the same rescaling is used as for
the uncertainty on the matrix element only (indeed, inclusion of the
PDF changes the best-fit rescaling by an amount which is essentially
irrelevant). 

Comparing Figure~\ref{fig:finres} with Figure~\ref{fig:orderdep}
shows again that
the Cacciari-Houdeau method, with rescaling when necessary,
provides an estimate of the theoretical uncertainty which is in
reasonable agreement with the behaviour of the perturbative expansion
at the known orders. This supports its use in order to estimate theoretical
uncertainties at the highest order at which they are known exactly
(NNLO) or approximately (N$^3$LO).  Figure~\ref{fig:finres} confirms
the conclusion we already reached by inspection of
Figure~\ref{fig:orderdep}, namely, that the dependence of results for
Higgs production in gluon fusion on the perturbative order of the PDF
is much weaker than that on the perturbative order of the matrix
element --- which, as well known, is unusually large. 
We conclude that an
exact determination of the N$^3$LO perturbative correction to the
matrix element will lead to a substantial reduction of the theoretical
uncertainty on the cross section for Higgs production in gluon fusion,
even without knowledge of N$^3$LO parton distributions.

\begin{figure}[t]
\centering
\begin{center}
\includegraphics[width=0.6\textwidth]{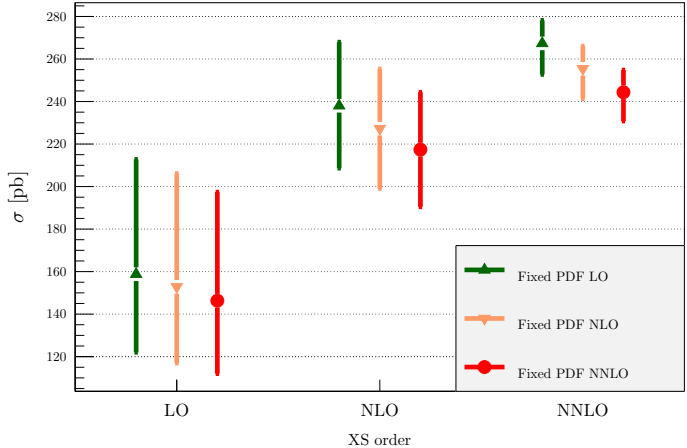} 
\end{center}
\caption{Same as Fig.~\ref{fig:orderdep}, but for top production.
The uncertainty bars are obtained by scale variation (see text).}
\label{fig:orderdeptop}
\end{figure}
While in the specific case of Higgs in gluon fusion the negligible
impact of N$^3$LO corrections to PDFs follows 
almost trivially from the huge hierarchy between the  uncertainty on
the PDF and that on the matrix element, 
one may ask whether this is true in general.

To see this, we have also considered the case of 
top pair production. The analogue of the plot of
Fig.~\ref{fig:orderdep} for the total top pair production
cross-section is shown in Fig.~\ref{fig:orderdeptop}. Results are
obtained using  {\tt TOP++2.0}~\cite{Czakon:2011xx}, including the recent full NNLO
result of Ref.~\cite{Czakon:2013goa}. Here too we take
$\alpha_s(M_Z)=0.119$, and we use  NNPDF2.3 PDFs (in the version with
maximum number of flavors $N_f=5$, as this is what  {\tt TOP++2.0}
requires).    
Uncertainty bars are now obtained by
varying both the renormalization and factorization scales by a factor
two about the central value $\mu_R=\mu_F=m_t$, with the ratio of the
two scales constrained not to exceed two~\cite{Czakon:2013tha}.

\begin{figure}[t]
\centering
\begin{center}
\includegraphics[width=0.6\textwidth]{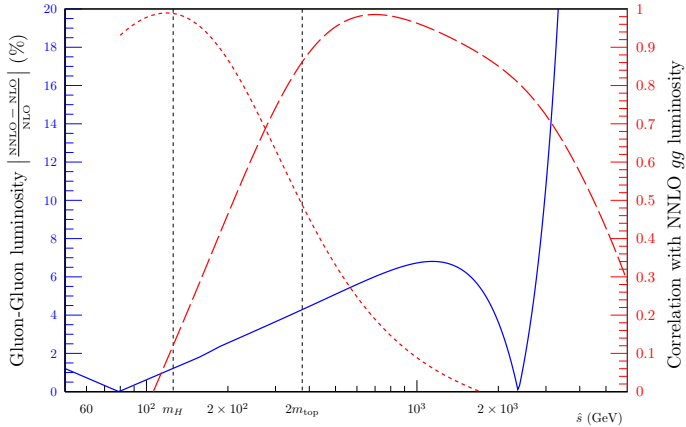} 
\end{center}
\caption{Percentage shift in gluon luminosity when going from NLO to
  NNLO (solid blue curve), compared to the correlation of the NNLO 
  gluon luminosity to the Higgs production cross section of
  Fig.~\ref{fig:orderdep} (red, short-dashed), and to the top
  production cross-section of Fig.~\ref{fig:orderdeptop} (red, long-dashed).}
\label{fig:correlation}
\end{figure}
It is
clear that, while the dependence on the order of the matrix element is
still somewhat stronger than that on the order of the PDF, now the two
are comparable, so that the dependence of the cross section on the
perturbative order with fixed PDF differs  by a non-negligible amount
from that found when the order of the PDF is
consistently varied along with that of the matrix element. In fact,
because the dependence on the order of the matrix element and that on
the order of the PDF are anti-correlated, the dependence of
the physical (hadron-level) cross-section on the perturbative order is
somewhat weaker than  found if the order of the matrix
element is varied while the PDF is kept fixed.

This example is sufficient to conclude that what is true for Higgs in
gluon fusion is not true in general: for other processes N$^3$LO
corrections to PDFs might well be relevant. Also,
the example raises two interesting questions. The first is the reason for
this difference between Higgs and top. The question can be answered by studying
 the perturbative behaviour of the
gluon luminosity, from which the dominant  contribution to both
processes originates, 
and comparing it to the correlation (defined as in 
Sect.~4 of~\cite{Alekhin:2011sk}) between the
gluon luminosity itself and the cross-sections which are being
computed, see Fig.~\ref{fig:correlation}. It is clear that the
correlation of the Higgs
cross section to the gluon luminosity is rather strongly peaked  in a
region in which the gluon luminosity depends very weakly on the
perturbative order, while the correlation of the top cross section is
large in a significantly broader kinematical  range (because even at LO
the invariant mass of the final state is not fixed), including a
region in which the perturbative dependence of the luminosity is
sizable. This implies that,
whereas the general behaviour of the cross-section 
may be easily understood in terms of the
features of the relevant physical processes and of the parton
luminosity (which in turn depends on the processes used for PDF
determination),  
whether the perturbative dependence of the PDF is or not important
has to be determined by a dedicated analysis of each process. In
particular, this requires a systematic correlation analysis, such as
that performed for several Higgs signal and background processes in
Sect.~3.2 of Ref.~\cite{Dittmaier:2012vm}; along with a study of the perturbative
dependence of parton luminosities.

The second question is how to best determine and use theoretical uncertainties
on PDFs. The perturbative behaviour of the top cross section of Fig.~\ref{fig:orderdeptop}
 suggests that the theoretical uncertainty on the top cross-section
 would be {\it over}estimated if the uncertainty on the PDF were not
 included, i.e. that the latter actually reduces the uncertainty of
 the physical cross section. Hence, in this case, in order to
 properly include the theoretical PDF uncertainty one must keep into
 account its (anti)correlation with the theoretical uncertainty on the
 matrix element. It appears that this would be  very difficult to do if theoretical PDF
 uncertainties were determined by scale variation. In this respect, a
 method such as Cacciari-Houdeau, based on the analysis of the
 perturbative behviour appears rather more promising. 

A systematic investigation of both these issues will be left for
further studies.

{\bf Acknowledgement:} we thank Alessandro Vicini for several
discussion and collaboration in the early stages of this
work. S.F. thanks Frank Petriello for asking the question in the title
of the paper, and Giampiero Passarino and Andr\'e David for
many enlightening discussions on theoretical uncertainties. This
research was supported in part by an Italian PRIN2009 grant, a
European Investment Bank  EIBURS grant, and by the European Commission through 
the ‘HiggsTools’ Initial Training
Network PITN-GA-2012-316704.


\begin{thebibliography}{99}
\baselineskip14pt


\bibitem{higgslo}
H.~M.~Georgi, S.~L.~Glashow, M.~E.~Machacek and D.~V.~Nanopoulos,
  Phys.\ Rev.\ Lett.\  {\bf 40} (1978) 692

\bibitem{higgsnlo}
  A.~Djouadi, M.~Spira and P.~M.~Zerwas,
  Phys.\ Lett.\ B {\bf 264} (1991) 440\\
 S.~Dawson,
  Nucl.\ Phys.\ B {\bf 359} (1991) 283\\
    M.~Spira, A.~Djouadi, D.~Graudenz and P.~M.~Zerwas,
  Nucl.\ Phys.\ B {\bf 453} (1995) 17
  [hep-ph/9504378]


\bibitem{higgsnnlo}
  R.~V.~Harlander and W.~B.~Kilgore,
  Phys.\ Rev.\ Lett.\  {\bf 88} (2002) 201801
  [hep-ph/0201206];\\
    C.~Anastasiou and K.~Melnikov,
  Nucl.\ Phys.\ B {\bf 646} (2002) 220
  [hep-ph/0207004];\\
 V.~Ravindran, J.~Smith and W.~L.~van Neerven,
  Nucl.\ Phys.\ B {\bf 665} (2003) 325
  [hep-ph/0302135].
\bibitem{Ball:2013bra}
 R.~D.~Ball, M.~Bonvini, S.~Forte, S.~Marzani and G.~Ridolfi,
  Nucl.\ Phys.\ B {\bf 874} (2013) 746
  [arXiv:1303.3590 [hep-ph]].
\bibitem{higgsn3lo}
 C.~Anastasiou, C.~Duhr, F.~Dulat and B.~Mistlberger,
  JHEP {\bf 1307} (2013) 003
  [arXiv:1302.4379 [hep-ph]];\\
 C.~Anastasiou, C.~Duhr, F.~Dulat, F.~Herzog and B.~Mistlberger,
  arXiv:1311.1425 [hep-ph];\\
  W.~B.~Kilgore,
  arXiv:1312.1296 [hep-ph].
\bibitem{Forte:2013wc}
  S.~Forte and G.~Watt,
  Ann.\ Rev.\ Nucl.\ Part.\ Sci.\  {\bf 63} (2013) 291
  [arXiv:1301.6754 [hep-ph]].
\bibitem{David:2013gaa}
  A.~David and G.~Passarino,
  Phys.\ Lett.\ B {\bf 726} (2013) 266
  [arXiv:1307.1843].
\bibitem{Cacciari:2011ze}
  M.~Cacciari and N.~Houdeau,
  JHEP {\bf 1109} (2011) 039
  [arXiv:1105.5152 [hep-ph]].
\bibitem{gghiggs} {\tt www.ge.infn.it/\textasciitilde bonvini/higgs}
\bibitem{Anastasiou:2011pi}
  C.~Anastasiou, S.~Buehler, F.~Herzog and A.~Lazopoulos,
  JHEP {\bf 1112} (2011) 058
  [arXiv:1107.0683 [hep-ph]];\\
{\tt http://www.phys.ethz.ch/\textasciitilde pheno/ihixs/}
\bibitem{Bonciani:2007ex}
  R.~Bonciani, G.~Degrassi and A.~Vicini,
  JHEP {\bf 0711} (2007) 095
  [arXiv:0709.4227 [hep-ph]].
\bibitem{Ball:2012cx}
  R.~D.~Ball, V.~Bertone, S.~Carrazza, C.~S.~Deans, L.~Del Debbio, S.~Forte, A.~Guffanti and N.~P.~Hartland {\it et al.},
  Nucl.\ Phys.\ B {\bf 867} (2013) 244
  [arXiv:1207.1303 [hep-ph]].
\bibitem{Ball:2011uy}
  R.~D.~Ball {\it et al.}  [NNPDF Collaboration],
  Nucl.\ Phys.\ B {\bf 855} (2012) 153
  [arXiv:1107.2652 [hep-ph]].
\bibitem{Carrazza:2013axa}
  S.~Carrazza, S.~Forte and J.~Rojo,
  arXiv:1311.5887 [hep-ph].
\bibitem{Olness:2009qd}
  F.~I.~Olness and D.~E.~Soper,
  Phys.\ Rev.\ D {\bf 81} (2010) 035018
  [arXiv:0907.5052 [hep-ph]].
\bibitem{Czakon:2011xx} 
  M.~Czakon and A.~Mitov,
  arXiv:1112.5675 [hep-ph].
\bibitem{Czakon:2013goa}
  M.~Czakon, P.~Fiedler and A.~Mitov,
  Phys.\ Rev.\ Lett.\  {\bf 110} (2013) 252004
  [arXiv:1303.6254 [hep-ph]].
\bibitem{Czakon:2013tha}
  M.~Czakon, M.~L.~Mangano, A.~Mitov and J.~Rojo,
  JHEP {\bf 1307} (2013) 167
  [arXiv:1303.7215 [hep-ph]].
\bibitem{Alekhin:2011sk}
  S.~Alekhin, S.~Alioli, R.~D.~Ball, V.~Bertone, J.~Blumlein, M.~Botje, J.~Butterworth and F.~Cerutti {\it et al.},
  arXiv:1101.0536 [hep-ph].
\bibitem{Dittmaier:2012vm}
  S.~Dittmaier, S.~Dittmaier, C.~Mariotti, G.~Passarino, R.~Tanaka, S.~Alekhin, J.~Alwall and E.~A.~Bagnaschi {\it et al.},
  arXiv:1201.3084 [hep-ph].
\end{thebibliography}
\end{document}